\documentclass[12pt]{article}

\usepackage{amsmath}
\usepackage{amssymb}
\usepackage{graphicx}
\usepackage{cite}
\usepackage{color} 
\usepackage{chemarrow}

\usepackage{setspace} 
\usepackage[labelfont=bf,labelsep=period,justification=raggedright]{caption}

\date{}

\begin{document}

\title{Collective Properties of a Transcription Initiation Model under Varying Environment}

\author{Yucheng Hu$^{1}$ and John S. Lowengrub$^{2, 3, 4, 5}$}

\maketitle

\begin{flushleft}
1. Zhou Pei-yuan center for applied mathematics, 
Tsinghua University, Beijing, China, 100008
\\
2. Department of Mathematics, University of California, Irvine, Irvine, CA 92697\\
3. Center for Complex Biological Systems, University of California, Irvine, Irvine CA 92697\\
4. Department of Biomedical Engineering, University of California, Irvine 92697\\
5. Chao Comprehensive Cancer Center, University of California, Irvine, Irvine, CA 92697
\\
$\ast$ E-mail: Corresponding huyc@tsinghua.edu.cn
\end{flushleft}

\begin{flushleft}
Mailing address\\

Yucheng Hu\\
116 Science Building, Tsinghua University, Beijing, China, 100008\\
Tel: 86-10-62794629, E-mail: huyc@tsinghua.edu.cn\\
John Lowengrub\\
540H Rowland Hall, University of California, Irvine, 92617\\
Tel: 1-(949)824-8456, E-mail: lowengrb@math.uci.edu\\
\end{flushleft}

\section*{Abstract}
The dynamics of gene transcription is tightly regulated in eukaryotes. Recent experiments have revealed various kinds of transcriptional dynamics, such as RNA polymerase II pausing, that involves regulation at the transcription initiation stage, and the choice of different regulation pattern is closely related to the physiological functions of the target gene. Here we consider a simplified model of transcription initiation, a process including the assembly of transcription complex and the pausing and releasing of the RNA polymerase II. Focusing on the collective behaviors on a population level, we explore potential regulatory functions this model can offer. These functions include fast and synchronized response to environmental change, or long-term memory about the transcriptional status. As a proof of concept we also show that, by selecting different control mechanisms cells can adapt to different environments. These findings may help us better understand the design principles of transcriptional regulation.


\section*{Introduction}

The capability to regulate behavior and physiological state in response to the environment is a fundamental property of all living systems. The transcription of most eukaryotic genes is subjected to primary regulation at the transcription level~\cite{darzacq2007vivo}. The transcription of a DNA template into messenger RNAs consists of a series of distinct phases including the binding of transcription factors to the DNA, the recruitment and sometimes pausing of RNA polymerase II (Pol II), the initiation and elongation of RNA transcripts, and eventually the termination of transcription~\cite{Core2008}. To understand the design principles of gene transcription regulation, it is important to identify which of these steps is rate-limiting and how this affects the transcription dynamics.

Traditional models often treat transcription as a two-state system, i.e., the activation and inactivation of target genes~\cite{Raj2006, Shahrezaei2008}. This is mostly true for prokaryotes. However, recent experiments using modern genetic techniques have provided  evidence of extensive transcriptional regulation acting on different stages during transcription process in eukaryotes~\cite{Sandelin2007,sun2011genetic}. For example, it is found that Pol II pausing is a general feature in mammalian cells, and the releasing of Pol II can be triggered by certain transcription factors such as c-Myc~\cite{Rahl2010, Young2011}. It has been argued that Pol II pausing is energetically costly, but doing so allows fast transcriptional response~\cite{boettiger2011transcriptional}. The willingness of cells to expend the extra energy in Pol II pausing suggests that fast transcriptional response of the target gene is important. Not surprisingly, genes that exhibit the Pol II pausing feature usually respond rapidly to signaling. The most studied example is the heatshock gene HSP90, whose Pol II pausing allows cells to act quickly (in minutes) in order to survive an abrupt temperature increase.

Another important feature some genes have is that they retain information about their previous transcriptional status despite environment changes. For example, it is found that knocking out of Brg in embryonic stem (ES) cells will eventually shut down the expression of the ES core transcription factors Oct4, Sox2 and Nanog, causing ES deficiency, but this process can take as long as several rounds of cell division after the knockout~\cite{Ho2009}. Persistent to noise and environmental fluctuations, this kind of self-sustained transcriptional activity is very useful during development, because it allows stem cells to maintain their identity.

The question of how the different transcriptional dynamics mentioned above are realized and encoded in the regulatory machinery has not been fully answered. Mathematical modeling can help elucidate these processes. Several mathematical models have been developed to understand the potential functions that transcriptional control can offer. In~\cite{rajala2010effects} the authors studied the effect of Pol II pausing on the transcriptional dynamics using a delayed stochastic model. Later, in~\cite{boettiger2011transcriptional} a detailed transcription regulatory model is presented, showing that regulation in Pol II pausing can lead to fast and synchronized transcription response. 

Continuing this line of work, we simplify and extend a transcription initiation model studied in~\cite{boettiger2011transcriptional} and study its transcriptional dynamics in a varying environment. Using a stochastic population model, we show that the collective transcriptional dynamics can be drastically different by letting the controlling signals to act on different steps of the transcription processes. In particular, we observe a fast and synchronized transcriptional response if the Pol II releasing step is under tight regulation, a result consistent with ~\cite{boettiger2011transcriptional}. We also discover a noise-resistant transcriptional dynamics that exhibits long-lasting memory effect, and to our best knowledge, this kind of behavior in transcriptional initiation has not been studied in detail before. Overall, this work highlights the importance of regulation during the transcription initiation process as a rich source of diversity in gene expression dynamics.

\section*{Results}
\subsection*{Transcription initiation model}
\begin{figure}[!ht]
\begin{center}
\includegraphics[width=.8\textwidth]{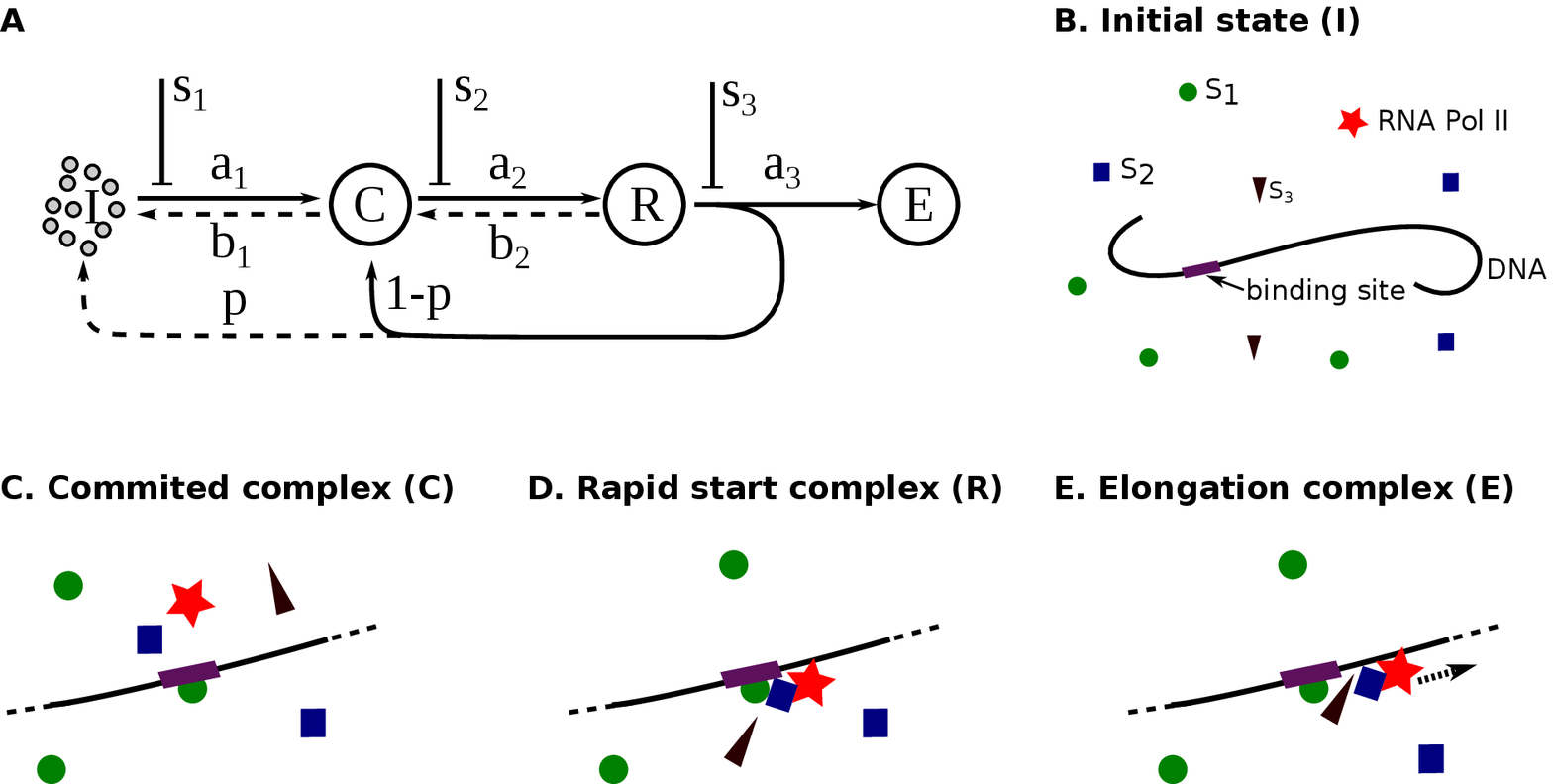}
\end{center}
\caption{
{\bf A gene transcription model with different controlling sites.}  
(A) In the model the transcription initiation process is divided into four distinct stages connected by reactions. The reaction rates $a_1$, $a_2$ and $a_3$ can be modified by controlling signal $s_1$, $s_2$ and $s_3$. The biological meaning of each stage is illustrated in (B-E). (B) Initial state: no transcription factors have bound to the DNA promoter and transcription initiation has not started yet. (C) Committed complex: a promoter (green circles) binds to the DNA binding site (purple bar). The committed complex is now ready to recruit other transcription factors and RNA Pol II. (D) Rapid start complex: the transcription machinery has finished assembly and is ready to release Pol II for transcription. (E) Elongated complex: released by the rapid start complex. The complex travels through the DNA template to carry out transcription. After releasing the elongated complex, the remaining part of the transcription complex can be reused as the committed complex for another round of transcription (solid loop in (A)). The transcript scaffold also has a small probability $p$ to be degraded in which case the transcription needs to start from the beginning (dashed loop in (A)).
}
\label{fig_model}
\end{figure}

We divide the transcriptional initiation process into four stages as depicted in Fig.~\ref{fig_model}. 
The initial state (I) represents the start of gene transcription in which transcription factors have not yet bound to DNA binding sites (Fig.~\ref{fig_model}B). After a promoter finds the specific DNA sequence (Fig.~\ref{fig_model}C) it will form a committed complex (C) which serves as a platform for other transcription factors and RNA Pol II to bind. A rapid start complex (R) is assembled when the Pol II is ready to be elongated for transcription (Fig.~\ref{fig_model}D). If certain conditions are met, the rapid start complex releases the elongated complex (E) and initializes the gene transcription (Fig.~\ref{fig_model}E). After ejecting the elongated complex, the remaining part of the rapid start complex on the DNA template, also referred to as the transcriptional scaffold, can function as a new committed complex. The recycling of the transcriptional scaffold makes gene transcription more efficient. With probability $p$, however, the scaffold will disassemble and for further transcription a new committed complex must be formed again from the initial state.

Advancement from one stage to the next is regulated by controlling signals, which act on three possible controlling sites, with signals $s_1, s_2$ and $s_3$ (Fig.~\ref{fig_model}A). In the following, they are referred to as the \emph{distal}, \emph{middle}, and \emph{proximal controlling site} according to their temporal distance to the final event of Pol II elongation. Biologically these signals may correspond to the downstream effects of rate-limiting proteins that participate in the transcription complex. Mathematically we assume $s_1$, $s_2$ and $s_3$ are variables ranging from 0 to 1, with 0 corresponding to the complete shutdown of the signal and 1 the fully opened state. We model the changes in the transcriptional state as stochastic chemical reactions. The forward reactions I $\rightarrow$ C, C $\rightarrow$ R, and R $\rightarrow$ E have reaction rates $a_1' = a_1 s_1$, $a_2' = a_2 s_2$ and $a_3' = a_3 s_3$, respectively. Here $a_1, a_2$ and $a_3$ are the intrinsic rates (when the signal $s_1 = s_2 = s_3 = 1$). The backwards reactions (C $\rightarrow$ I and R $\rightarrow$ C) have fixed reaction rates $b_1$ and $b_2$. After elongation, part of the transcription complex, including RNA Pol II, will leave the rapid start complex. Whether the scaffold will be reused or discarded is determined by probabilities $1-p$ and $p$, respectively. We assume the controlling sites are independent of each other. See Materials and Methods for a more detailed description of the model.

All together there are 6 parameters in the model and their values are important for the transcription dynamics. In the following we assume that (i) $a_1 \ll a_2 \ll a_3$, (ii) $b_1 \ll a_1$, $b_2 \ll a_2$, and (iii) $p$ is small. The heuristics behind assumption (i) is that as the regulation becomes more precise and specific at the later stages of the transcription initiation, the transcription machinery may want to speed up the process by using more energy. Note that assumption (iii) is essential for the long term memory effect in transcription (see below). In our simulation, we use $a_1 = 0.1, a_2 = 1, a_3 = 10$, $b_1 = 0.02, b_2 = 0.2 \ (\mathrm{hour}^{-1})$ and $p=0.1$. According to this setting, when $s_1 = s_2 = s_3 = 1$, on average it takes about 10 hours from the initial state to form a committed complex; 1 hour from the committed complex to the rapid star complex; and 6 minutes for the rapid start complex to release an elongated complex. These values are consistent, at least in order of magnitude, with \emph{in vitro} observations of the transcription dynamics~\cite{Hawley1985} (here \emph{in vitro} data may be better than \emph{in vivo} data as the former reflects the intrinsic rates $a_1, a_2, a_3$ rather than the regulated rates $a_1', a_2', a_3'$). As shown in~\cite{boettiger2011transcriptional}, there may be many choices of reaction rate parameters that lead to similar behaviors. Here our main purpose is to study the transcription dynamics by letting the regulatory signal acting on different controlling sites under the same model parameters.

\subsection*{Static properties of the model}
\begin{figure}[!ht]
\begin{center}
\includegraphics[width=4in]{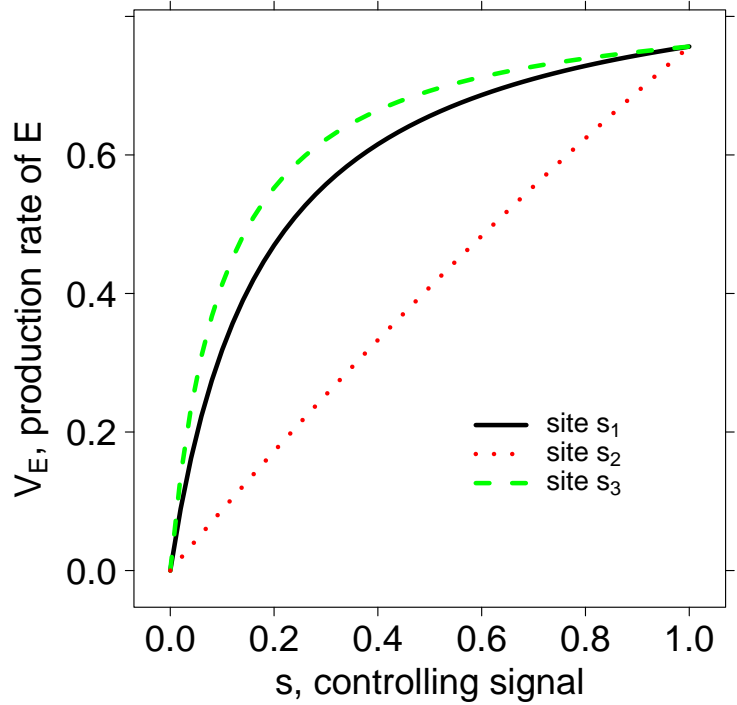}
\end{center}
\caption{
{\bf The dependence of the production rate of E on the controlling signal.}
For a fixed controlling signal $s$ whose value ranges from 0 to 1, we determine $V_E^i(s)$, the production rates of $E$,  by taking $s_i = s$ for one $i=1,2,3$ while keeping the other two signals $s_j=1 (i\ne j)$ in the ON-state. The equilibrium production rates of $E$, given by Eqs.~\eqref{eq_eplevel1}-\eqref{eq_epleve3} are plotted here.}
\label{fig_Eprod}
\end{figure}

First we change the signal strengths $s_1$, $s_2$ and $s_3$ at the distal, middle and proximal controlling sites, respectively, to see how the transcription rate at equilibrium responds. For each controlling site, say the distal one, we keep the other two open ($s_2 = s_3 = 1$) and let $s_1$ vary from 0 to 1. For a fixed $s_i, i=1,2,3$, the transcription rate is measured by the rate of production of the elongated complex, $V_E^i=a_3' P_R$, where $P_R$ is the probability the system being at state R. At equilibrium, the in-flux and out-flux of each state should be equal and in this case $P_R$ can be solved explicitly (see the Materials and Methods for details). As a result, we obtain
\begin{eqnarray}
 V_E^1(s) &=& \frac{s a_1 a_3}{\frac{b_2+a_3}{a_2} b_1 + a_3p + \left( \frac{a_2+b_2+a_3}{a_2}\right) s a_1},  \label{eq_eplevel1} \\
 V_E^2(s) &=& \frac{s a_2 a_3}{b_2 + a_3 + \frac{b_2+a_3}{a_1}b_1 + \frac{a_1 + a_3p}{a_1}s a_2},  \label{eq_eplevel2} \\
 V_E^3(s) &=& \frac{s a_2 a_3}{\frac{a_1+b_1}{a_1} b_2 + a_2 + \left( \frac{a_1+b_1+a_2p}{a_1}\right) s a_3}.  \label{eq_epleve3}
\end{eqnarray}
We can see that, the dependence of the transcription rate on the signal strength $s$ takes the form of a Hill's function for all the three control mechanisms. However, the exact shapes of the functions are quite different (see Fig.~\ref{fig_Eprod}). The transcription rate is almost linearly dependent on the signal strength at the middle controlling site but for the other two the rates quickly saturate as the signal strength increases.

For the proximal controlling site $s_3$, because the intrinsic rate $a_3$ is very large, a very small $s_3$ makes this step of reaction (P $\rightarrow$ E) rate-limiting. As a result, the transcription rate, which is proportional to $a_3 s_3$ in this limit, is very sensitive to changes in $s_3$. However, as $s_3$ increases, the upstream reaction cannot supply enough rapid start complex so the transcription rate saturates. For the distal controlling site $s_1$, because of the recycling of the transcription scaffold, a small production of the committed complex is enough to compensate for the loss of the committed complex caused by the degradation of the scaffold ($p$ needs to be small). As $s_1$ keeps increasing, the downstream reaction reach their limit and the transcription rate saturates.  

\subsection*{Dynamical properties of the model}
For all living systems the ability to adjust their gene expression in response to developmental signals and environmental changes is crucial for their survival. In the following, we focus on the dynamical properties of the model when it respond to time-varying signals.

\subsubsection*{Pol II pausing and synchronized expression}
First we introduce a regulatory signal at the proximal controlling site $s_3$, which switches between 1 and 0 periodically every 12 hours, while the other two controlling sites are in the ON-state throughout the simulation. We simulate a population of 8,000 cells and measure the instantaneous transcription rate, which is the total amount of the production of the elongated complex within a short time interval. Initially all cells are at transcription state I. During the first cycle in which the signal at $s_3$ is turned on, the upstream reactions (I $\rightarrow$ C and C $\rightarrow$ R) are the rate-limiting steps, and the expression level gradually increases from zero to a steady state set by the maximum rate determined by the upstream reactions and the value of $p$ (Fig.~\ref{fig_signal} top panel). When the signal at $s_3$ is set to OFF, the upstream reactions will keep working to produce rapid start complex, which can be recycled or degraded back to state C. Because $b_2 \ll a_2$, most cells will stay at state R with their Pol II at a poised state waiting to be launched. Now if the signal at $s_3$ is turned on, the paused Pol II is released in a very short period of time, causing a transient burst in the transcription level. However, this fast transcription rate can not last because the upstream reactions have a limited supply. As a result, the transcription quickly drops back to the equilibrium level.

If the same signal is acting on the middle controlling site $s_2$ instead of $s_3$, the overall expression level will drop to zero as the signal is turned off and increase to a moderate level as the signal is turned on. Because the accumulation of committed complex formed during the signal-OFF period, the expression level reaches a small peak after the signal is turned on. However, compared with the case of regulating the proximal site, the response time is relatively longer, and the peak expression level is much smaller. In other words, cells independently release their Pol II after the rapid start complex is formed, and there is little synchronized expression in the population.

\subsubsection*{Memory effect and noise suppression}
Interestingly, when the regulating signal acts on the distal controlling site $s_1$, the transcription of the target gene will last for a long period of time even after the signal is turned off (Fig.~\ref{fig_signal} bottom panel). This is because the transcription scaffold can be reused as a committed complex for many times without the need to assemble a new one from the beginning. The time that the self-sustained transcription persists depends on the value of $p$. For very small $p$ (very stable transcription scaffold), the expression can last for days when the activation signal is gone. This kind of transcriptional behavior may explain the memory effect found in certain genes discussed earlier.

\begin{figure}[!ht]
\begin{center}
\includegraphics[width=.8\textwidth]{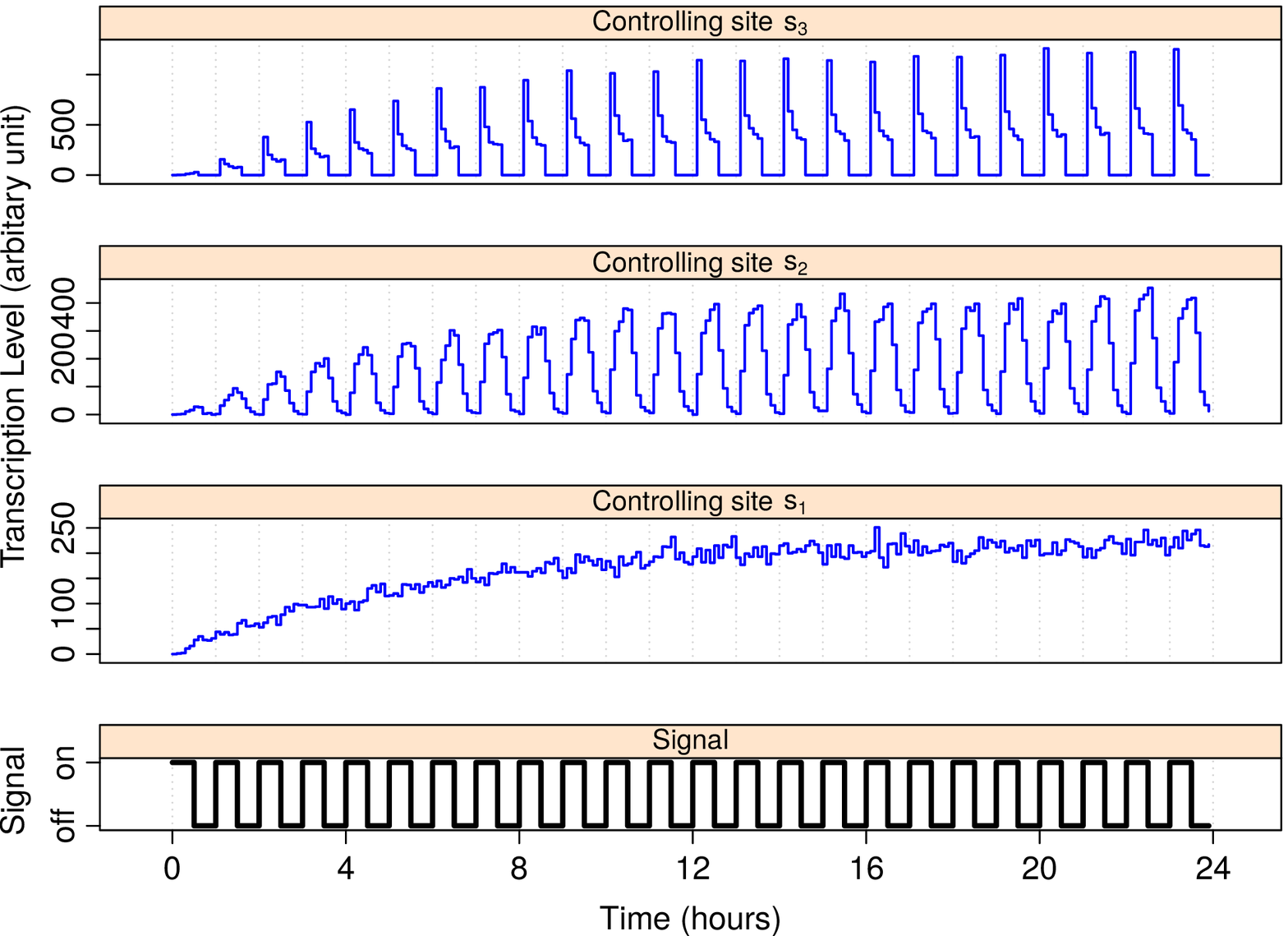}
\end{center}
\caption{
{\bf Collective behavior of gene transcription under periodic signal (short period).}
A population of 8,000 cells is simulated over time, during which we apply an oscillating signal ($T=2$ hours, bottom panel) to one of the three controlling sites of of the transcription module embedded in the cells. At time t=0, all cells are at the initial state. The transcription level is obtained by summing the total production of $E$ in the population during a short period of time (0.1 hour).}
\label{fig_signal_fast}
\end{figure}

Next we apply a high frequency signal (0.5 hour ON followed by 0.5 hour OFF) at the three controlling sites respectively. As Fig.~\ref{fig_signal_fast} shows, for the distal controlling site $s_1$, the transcription level maintains a steady state even though the signal is changing. For the other two controlling site $s_2$ and $s_3$, however, the signal change causes significant oscillations during transcription. This result shows that it is possible for the transcription machinery to suppress high frequency noise in the signal.

\subsection*{Evolvability of the transcription module}
We have shown that different transcriptional regulation can give rise to different expression dynamics. Next we demonstrate that environmental conditions can direct adaptation of transcriptional regulation in a population of cells. We consider three types of cells whose gene of interest is under different transcription control: the target gene of the type-1 cells is under distal control (signal acting on $s_1$ while $s_2$ and $s_3$ are constantly ON); type-2 cells under middle control (signal acting on $s_2$ while $s_1$ and $s_3$ are constantly ON) and type-3 cells under proximal control (signal acting on $s_3$ while $s_1$ and $s_2$ are constantly ON). Here the signal represents the level of growth-promoting factors (GPFs) in the environment, which alternates between ON (high level of GPFs) and OFF (low level of GPFs) with a given period $T$. This setup mimics the experimental strategy used in~\cite{Poelwijk2011} in which a genetic module is engineered whose expression is beneficial in one environment condition and detrimental in another. In~\cite{Poelwijk2011}, a variable environment that switches every 6 hours between beneficial and detrimental conditions is used to examine the evolution of this genetic module.

Here we use the Moran population model~\cite{Moran1958} in which cells can either replicate or die and investigate the dynamics of the system under competition and selective pressure. The fitness of each cell is modeled by an auxiliary variable called $F$, which is determined by the following rules: when GPFs are abundant (ON signal), successfully producing an elongated complex increases the $F$ by a certain amount; when GPFs are low (OFF signal), producing an elongated complex does not change $F$. Meanwhile, every time the transcriptional state advances to the next stage, $F$ decreases (here we assume only the forward reactions cost energy while the backward ones do not). When the $F$ level of a cell drops below a critical value, transcription activity is paused. When $F$ reaches a certain threshold the cell divides into two daughter cells. $F$ in the mother cells is equally splitted into the two daughter cells which inherit the same regulation mechanism with their mother cell. When a cell divides, one cell in the population will be chosen to be replaced according to their $F$ level: the chance of being chosen is proportional to the inverse of $F$. This means cells with smaller $F$ are more likely to die. See Appendix for more details about the model, the parameters and implementation details.

\begin{figure}[!ht]
\begin{center}
\includegraphics[width=.8\textwidth]{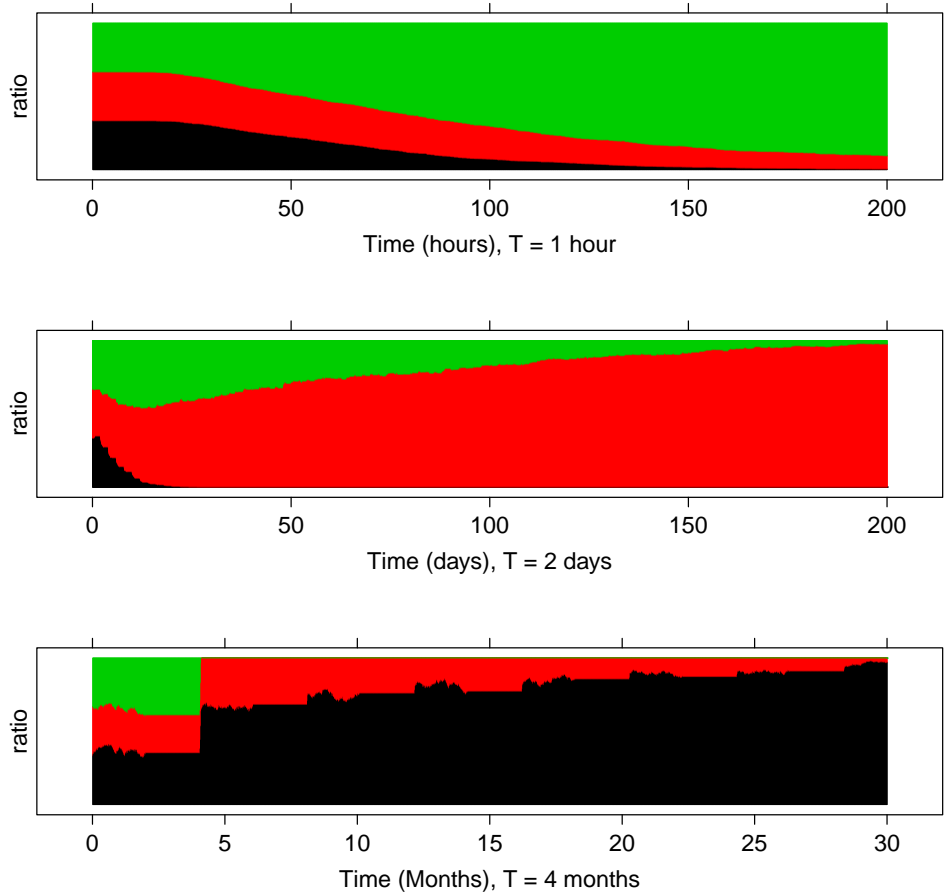}
\end{center}
\caption{
{\bf Evolution of a heterogeneous population over time.} 
The population (2,000 cells in total) is composed of type-1, type-2 and type-3 cells, which are distinguished by the way the transcription initiation is regulated (see the main text). Initially the proportions of the three cell types are equal. Over time the proportion of each type is indicated by the thickness of the colored layer (Black: distal control, type-1; Red: middle control, type-2; Green: proximal control, type-3). A growth promoting signal is switched between ON and OFF periodically in time with period $T$. Each panel corresponds to a different $T$ (Top: $T=1$ hour, Middle: $T=2$ days, Bottom: $T=4$ months). The results show that different regulatory mechanisms have different fitness depending on periodicity of the signal: for fast-switching signals type-1 cells are dominant while for slow-switching signals type-3 cells become the most dominant. In between, type-2 cells are the most fit.}
\label{fig_evolution}
\end{figure}

The numerical results show that the performance of the three types of transcription mechanisms depends on the period $T$ at which the environment changes. The one that best fits the environment will become dominant eventually. As Fig.~\ref{fig_evolution} shows, under a rapidly changing environment (small $T$), type-3 cells are the most fit type because they can respond quickly to rapid changes in GPFs (Fig.~\ref{fig_evolution}A, black: type-1 cell; red: type-2 cell; green: type-3 cell). The energy that type-3 cells spend at Pol II pausing pays off in this situation. However, as $T$ increases, type-3 cells keep consuming energy during the long period of low GPFs, which decreases their fitness. In contrast, the other two cell-types do a better job in preserving energy during the OFF state. As a result, for $T=2$ days, type-2 cells are most fit (Fig.~\ref{fig_evolution}B), and for even longer periods ($T = 4$ months), the type-1 cells dominate the population(Fig.~\ref{fig_evolution}C). Through this game of life we show that, transcription regulation in a population of cells can adapt to meet different regulatory demands. 

\section*{Discussion}
Recent experimental results on Pol II pausing demonstrate that transcriptional regulation is widespread and important in eukaryotes~\cite{darzacq2007vivo}. Nevertheless, the design principles behind the different kinds of transcriptional dynamics have not been fully understood. To aid in closing the gap, we developed a stochastic transcription model focusing on the regulatory steps during the assembly of the transcription complex and Pol II pausing. Our simulation results showed that different control mechanisms can lead to distinct collective transcription behaviors, which offers great flexibility in regulating the transcription of the target genes. 

\begin{figure}[!ht]
\begin{center}
\includegraphics[width=.8\textwidth]{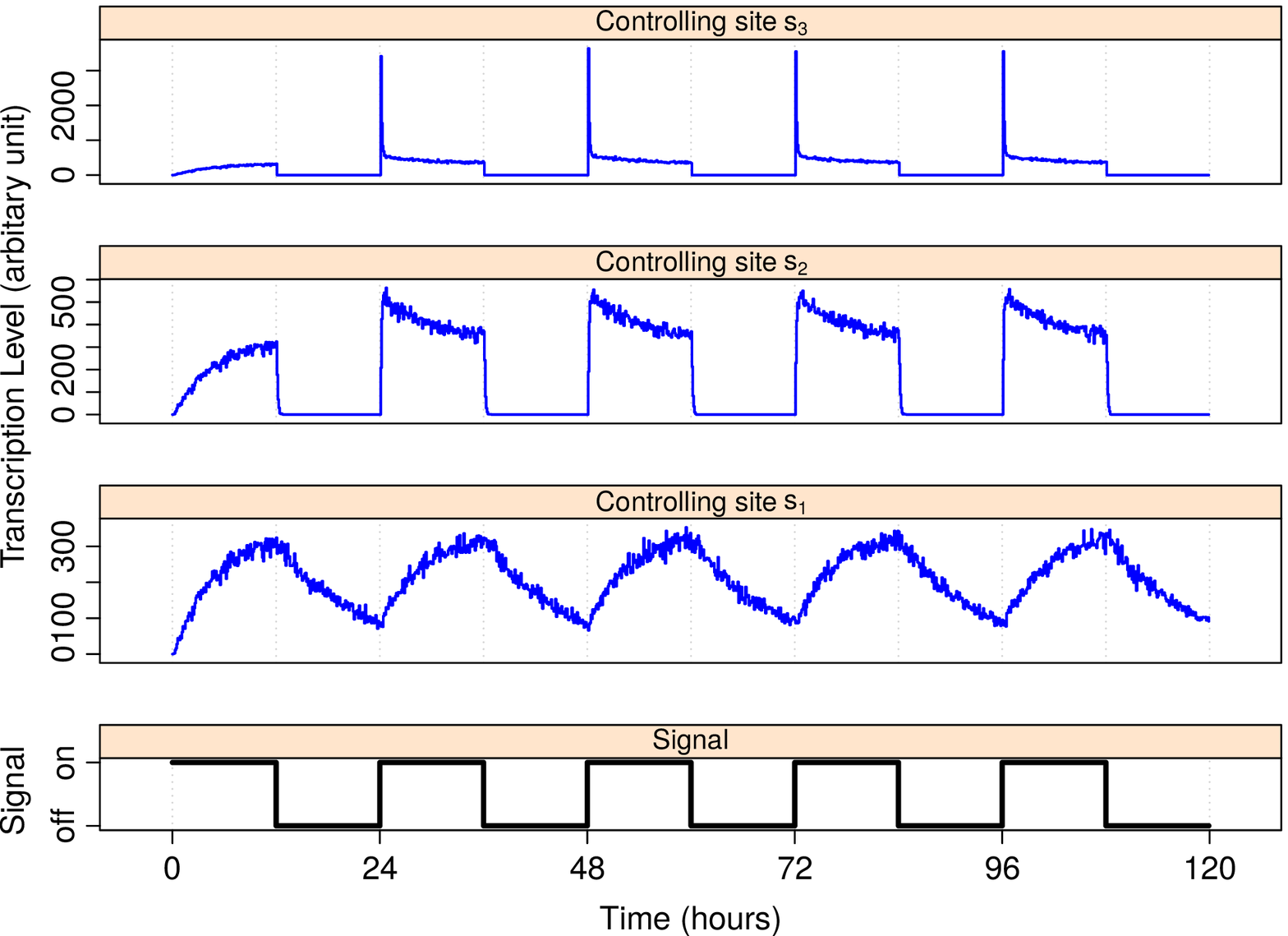}
\end{center}
\caption{
{\bf Collective behavior of gene transcription under periodic signal (long period).}
A population of 8,000 cells is simulated over time, during which we apply an oscillating signal ($T=24$ hours, bottom panel) to one of the three controlling sites of the transcription module which is embedded in all the cells. At time t=0, all cells are at the initial state. The transcription level is obtained by summing the total production of $E$ in the population during a short period of time (0.1 hour).}
\label{fig_signal}
\end{figure}

On the one hand, regulations right before transcription such as Pol II pausing (corresponds to controlling the proximal site $s_1$ in our model) can make gene expression fast and synchronized (Fig.~\ref{fig_signal} top panel). This kind of expression dynamics allows cells to rapidly respond to environmental changes or regulatory signals. Indeed, recent experiments have shown that Pol II pausing at promoter-proximal site of many genes tends to respond rapidly to developmental and cell signaling in mammals~\cite{Core2008}.

On the other hand, regulations early on at the transcription initiation processes (corresponds to controlling the distal site $s_3$ in our model) may give rise to noise-resistant and self-sustained expression pattern (Fig.~\ref{fig_signal} and~\ref{fig_signal_fast}). We suspect that the recycling of the transcription scaffold is a way of retaining gene transcription information. The ability for transcription to persist in the presence of fluctuating signals are very important for developmental purposes. For example, it may help pluripotent cells to maintain their stem-cell identity. Because differentiation is mostly an irreversible process during development, fate decisions need to be made with caution. In the case of Brg in ESCs we mentioned earlier~\cite{Ho2009}, it is possible that Brg is a signal regulating the expression of ES core transcription factors whose role is similar to the distal controlling signal $s_1$. As a result, those ES core transcription factors have a long-term self-sustained expression pattern mimicked by Fig.~\ref{fig_signal} (``Controlling site $s_1$'').

We showed that rich expression patterns can be achieved under our simple transcription initiation model. Thus it is possible that the transcriptional machinery could be adapted by different genes to fullfil different biological functions. Using a game of life, we showed that such evolvability is important for cells to survive in changing environments. We hope these functional analysis for the simplified transcription initiation model may help us understand the logic of transcriptional control.

\section*{Materials and Methods}
\subsection*{Model}
The model in Fig.~\ref{fig_model} can be described by a continuous-time Markov process with states I, C, R connected by the following reaction rules:
\begin{eqnarray*}
 I &\autorightleftharpoons{$s_1 a_1$}{$b_1$}& C, \\
 C &\autorightleftharpoons{$s_2 a_2$}{$b_2$}& R, \\
 R &\autorightarrow{$s_3 a_3 (1-p)$}{}& C + E, \\
 R &\autorightarrow{$s_3 a_3 p$}{}& E.
\end{eqnarray*}
This system can be simulated using Gillespie's direct method~\cite{Gillespie1977}. The last two reactions produce an elongated complex. The total number of elongated complexes of all cells in the population within every six minutes is recorded and is used to measure the instantaneous transcription rate, which is the value on the y-axis in Figs.~\ref{fig_signal} and~\ref{fig_signal_fast}.

\subsection*{Equilibrium transcription rate}
The transcription rate at equilibrium as a function of the strength of the controlling signal can be obtained analytically as follows. At any given time, the system must be at one of the three possible states, I, C or R. Let $P_I$, $P_C$ and $P_R$ be the probabilities that the system is in state I, C, and R, respectively. Consequently,
\begin{equation}
P_I + P_C + P_R = 1.
\label{eq_probconsv}
\end{equation}
At equilibrium, the system must have reached detailed balance, i.e., the flux into each state must be equal to flux out of each state, which gives
\begin{eqnarray*}
P_C b_1 + P_R a_3' p &=& P_I a_1', \\
P_I a_1' + P_R b_2 + P_R a_3' (1-p) &=& P_C b_1 + P_C a_2', \\
P_C a_2' &=& P_R a_3' + P_R b_2.
\end{eqnarray*}
Here $a_1' = s_1 a_1$, $a_2' = s_2 a_2$, and $a_3' = s_3 a_3$ are the regulated rate under the controlling signal. The production rate of E is proportional to $P_R$ times the reaction rate from R to E, that is,
\begin{equation}
V_E = P_R a_3'.
\label{eq_Eprod_0}
\end{equation}
Solving the above equations we obtain the equilibrium transcription rates in Eqs.~\eqref{eq_eplevel1}-\eqref{eq_epleve3}.

\section*{Acknowledgments}
YH is supported by the NSFC (National Science Foundation of China) under Grant No.~11301294.

\section*{Appendix: Implementation details of the evolutionary model}
\noindent\textbf{Gene transcription}
The gene of interest of each cell in the population is regulated by external signal (environment) acting on one of the three controlling sites of the transcription module. This gives us type-1, type-2 and type-3 cells which regulated by controlling site $s_1$, $s_2$ and $s_3$, respectively. The signal periodically switches between 1 (ON signal) and 0 (OFF signal) with period $T$ (a duration of $T/2$ ON followed by $T/2$ OFF, and so on). During the evolution, we monitor the fitness $F$ for each cell. The three forward reactions (I $\rightarrow$ C, C $\rightarrow$ R and R $\rightarrow$ E) consume two units of $F$ of the host cell each time they fire. Backwards reactions (C $\rightarrow$ I and R $\rightarrow$ C) do not decrease or increase $F$. After a successful transcription, an E will be produced, whose effect on $F$ depends on the current environment: if the environment is at the ON-state when E is made, ten units will be added to $F$; however, if the environment is at the OFF-state, $F$ does not change. The rationale behind these choices is that, when there are enough GPFs, the subsequent transcription and translation processes can be successfully carried out and help the cell to gain more GPFs; when there are not enough GPFs, these processes may get stuck or aborted, which does not provide any benefit to the cell. Note that this model of environment regulation of transcription is highly simplified and ignores post-transcriptional processes, which may be highly complex.\\

\noindent\textbf{Cell growth} 
Initially each cell in the population has 50 $F$ units. As soon as a cell acquires 100 points it will divide, and its $F$ units will be equally divided into the two daughter cells. We assume the population evolves according to the Moran model~\cite{Moran1958}. When a cell divides, another randomly chosen cell in the population will be removed. The probability that a cell is chosen to die is proportional to the inverse of the value of $F$ that cell currently has. As a result, the total population (2000 cells) will remain constant during evolution.

After replication, the two daughter cells will inherit the same regulation mechanism as the mother cell. In order to mimic the epigenetic effect, one of the daughter cells will share the same transcriptional state (I, C, or R) of its mother cell while the other is set to initial state. We also tested other implementations, such as both daughter cells start with the initial state, or both inherit the transcriptional state of their mother cell. The results are similar.

According to the above setting, when the GPFs are abundant (signal constantly ON), the average cell division cycle is around 25 hours, which is reasonable for mammalian cells.

\bibliographystyle{siam}
\bibliography{ref}

\newpage
\section*{Figure Legends}

%

\end{document}